%
%
%
%
%
%
%
%
\def\standardrisposta{s }\def\reducedrisposta{r }
\def\mplarisposta{mpla }\def\zerorisposta{z }
\def\doublerisposta{d }\def\cartarisposta{e }\def\amsrisposta{y }
\newcount\ingrandimento \newcount\sinnota \newcount\dimnota
\newcount\unoduecol \newdimen\collhsize \newdimen\tothsize
\newdimen\fullhsize \newcount\controllorisposta \sinnota=1
\newskip\infralinea  \global\controllorisposta=0
%
%
%
%
\def\risposta{s }
\def\srisposta{e }
\def\arisposta{y }
\ifx\risposta\standardrisposta \ingrandimento=1200
\message {>> This will come out UNREDUCED << }
\dimnota=2 \unoduecol=1 \global\controllorisposta=1 \fi
\ifx\risposta\reducedrisposta \ingrandimento=1095 \dimnota=1
\unoduecol=1  \global\controllorisposta=1
\message {>> This will come out REDUCED << } \fi
\ifx\risposta\doublerisposta \ingrandimento=1000 \dimnota=2
\unoduecol=2   \message {>> You must print this in
LANDSCAPE orientation << } \global\controllorisposta=1 \fi
\ifx\risposta\mplarisposta \ingrandimento=1000 \dimnota=1
\message {>> Mod. Phys. Lett. A format << }
\unoduecol=1 \global\controllorisposta=1 \fi
\ifx\risposta\zerorisposta \ingrandimento=1000 \dimnota=2
\message {>> Zero Magnification format << }
\unoduecol=1 \global\controllorisposta=1 \fi
\ifnum\controllorisposta=0  \ingrandimento=1200
\message {>>> ERROR IN INPUT, I ASSUME STANDARD
UNREDUCED FORMAT <<< }  \dimnota=2 \unoduecol=1 \fi
\magnification=\ingrandimento
%
%
%
%
\newdimen\eucolumnsize \newdimen\eudoublehsize \newdimen\eudoublevsize
\newdimen\uscolumnsize \newdimen\usdoublehsize \newdimen\usdoublevsize
\newdimen\eusinglehsize \newdimen\eusinglevsize \newdimen\ussinglehsize
\newskip\standardbaselineskip \newdimen\ussinglevsize
\newskip\reducedbaselineskip \newskip\doublebaselineskip
\eucolumnsize=12.0truecm    
\eudoublehsize=25.5truecm   
\eudoublevsize=6.7truein    
\uscolumnsize=4.4truein     
\usdoublehsize=9.4truein    
\usdoublevsize=6.8truein    
\eusinglehsize=6.5truein    
\eusinglevsize=24truecm     
\ussinglehsize=6.5truein    
\ussinglevsize=8.9truein    
\standardbaselineskip=16pt plus.2pt  
\reducedbaselineskip=14pt plus.2pt   
\doublebaselineskip=12pt plus.2pt    
%
%
\def\Portoffset{}
\def\Landoffset{\voffset=-.2truein}
\ifx\risposta\mplarisposta \def\Portoffset{\hoffset=1.8truecm} \fi
%
%
\def\Landspec{}
\tolerance=10000
\parskip=0pt plus2pt  \leftskip=0pt \rightskip=0pt
%
%
\ifx\risposta\standardrisposta \infralinea=\standardbaselineskip \fi
\ifx\risposta\reducedrisposta  \infralinea=\reducedbaselineskip \fi
\ifx\risposta\doublerisposta   \infralinea=\doublebaselineskip \fi
\ifx\risposta\mplarisposta     \infralinea=13pt \fi
\ifx\risposta\zerorisposta     \infralinea=12pt plus.2pt\fi
\ifnum\controllorisposta=0    \infralinea=\standardbaselineskip \fi
\ifx\risposta\doublerisposta   \Landoffset \else \Portoffset \fi
\ifx\risposta\doublerisposta \ifx\srisposta\cartarisposta
\tothsize=\eudoublehsize \collhsize=\eucolumnsize
\vsize=\eudoublevsize  \else  \tothsize=\usdoublehsize
\collhsize=\uscolumnsize \vsize=\usdoublevsize \fi \else
\ifx\srisposta\cartarisposta \tothsize=\eusinglehsize
\vsize=\eusinglevsize \else  \tothsize=\ussinglehsize
\vsize=\ussinglevsize \fi \collhsize=4.4truein \fi
\ifx\risposta\mplarisposta \tothsize=5.0truein
\vsize=7.8truein \collhsize=4.4truein \fi
%
%
%
%
\newcount\contaeuler \newcount\contacyrill \newcount\contaams
\font\ninerm=cmr9  \font\eightrm=cmr8  \font\sixrm=cmr6
\font\ninei=cmmi9  \font\eighti=cmmi8  \font\sixi=cmmi6
\font\ninesy=cmsy9  \font\eightsy=cmsy8  \font\sixsy=cmsy6
\font\ninebf=cmbx9  \font\eightbf=cmbx8  \font\sixbf=cmbx6
\font\ninett=cmtt9  \font\eighttt=cmtt8  \font\nineit=cmti9
\font\eightit=cmti8 \font\ninesl=cmsl9  \font\eightsl=cmsl8
\skewchar\ninei='177 \skewchar\eighti='177 \skewchar\sixi='177
\skewchar\ninesy='60 \skewchar\eightsy='60 \skewchar\sixsy='60
\hyphenchar\ninett=-1 \hyphenchar\eighttt=-1 \hyphenchar\tentt=-1
\def\bfmath{\cmmib}                 
\font\tencmmib=cmmib10  \newfam\cmmibfam  \skewchar\tencmmib='177
\font\tencmbsy=cmbsy10  \newfam\cmbsyfam  \skewchar\tencmbsy='60
\def\scaps{\cmcsc}                 
\font\tencmcsc=cmcsc10  \newfam\cmcscfam
\ifnum\ingrandimento=1095

\font\capsone=cmcsc10 at 10.95pt 

\else

\font\capsone=cmcsc10 at 12pt 
\fi

\def\ttaarr{\bf}		
\def\ppaarr{\sl}		

%
%
%
\newfam\eufmfam \newfam\msamfam \newfam\msbmfam \newfam\eufbfam
\def\Loadeulerfonts{\global\contaeuler=1 \ifx\arisposta\amsrisposta
\font\teneufm=eufm10              
\font\eighteufm=eufm8 \font\nineeufm=eufm9 \font\sixeufm=eufm6
\font\seveneufm=eufm7  \font\fiveeufm=eufm5
\font\teneufb=eufb10              
\font\eighteufb=eufb8 \font\nineeufb=eufb9 \font\sixeufb=eufb6
\font\seveneufb=eufb7  \font\fiveeufb=eufb5
\font\teneurm=eurm10              
\font\eighteurm=eurm8 \font\nineeurm=eurm9
\font\teneurb=eurb10              
\font\eighteurb=eurb8 \font\nineeurb=eurb9
\font\teneusm=eusm10              
\font\eighteusm=eusm8 \font\nineeusm=eusm9
\font\teneusb=eusb10              
\font\eighteusb=eusb8 \font\nineeusb=eusb9
\else \def\eufm{\tt} \def\eufb{\tt} \def\eurm{\tt} \def\eurb{\tt}
\def\eusm{\tt} \def\eusb{\tt}    \fi}
\def\loadeuler{\Loadeulerfonts\tenpoint}
\def\loadamsmath{\global\contaams=1 \ifx\arisposta\amsrisposta
\font\tenmsam=msam10 \font\ninemsam=msam9 \font\eightmsam=msam8
\font\sevenmsam=msam7 \font\sixmsam=msam6 \font\fivemsam=msam5
\font\tenmsbm=msbm10 \font\ninemsbm=msbm9 \font\eightmsbm=msbm8
\font\sevenmsbm=msbm7 \font\sixmsbm=msbm6 \font\fivemsbm=msbm5
\else \def\msbm{\bf} \fi \def\Bbb{\msbm} \def\symbl{\msam} \tenpoint}
\def\loadcyrill{\global\contacyrill=1 \ifx\arisposta\amsrisposta
\font\tenwncyr=wncyr10 \font\ninewncyr=wncyr9 \font\eightwncyr=wncyr8
\font\tenwncyb=wncyr10 \font\ninewncyb=wncyr9 \font\eightwncyb=wncyr8
\font\tenwncyi=wncyr10 \font\ninewncyi=wncyr9 \font\eightwncyi=wncyr8
\else \def\cyrill{\sl} \def\cyrilb{\sl} \def\cyrili{\sl} \fi\tenpoint}
\ifx\arisposta\amsrisposta
\font\sevenex=cmex7               
\font\eightex=cmex8  \font\nineex=cmex9
\font\ninecmmib=cmmib9   \font\eightcmmib=cmmib8
\font\sevencmmib=cmmib7 \font\sixcmmib=cmmib6
\font\fivecmmib=cmmib5   \skewchar\ninecmmib='177
\skewchar\eightcmmib='177  \skewchar\sevencmmib='177
\skewchar\sixcmmib='177   \skewchar\fivecmmib='177
%
%
%
\def\ninecmbsy{\tencmbsy}
\def\eightcmbsy{\tencmbsy}
\def\sevencmbsy{\tencmbsy}
\def\sixcmbsy{\tencmbsy}
\def\fivecmbsy{\tencmbsy}
\font\ninecmcsc=cmcsc9    \font\eightcmcsc=cmcsc8     \else
\def\cmmib{\fam\cmmibfam\tencmmib}\textfont\cmmibfam=\tencmmib
\scriptfont\cmmibfam=\tencmmib \scriptscriptfont\cmmibfam=\tencmmib
\def\cmbsy{\fam\cmbsyfam\tencmbsy} \textfont\cmbsyfam=\tencmbsy
\scriptfont\cmbsyfam=\tencmbsy \scriptscriptfont\cmbsyfam=\tencmbsy
\scriptfont\cmcscfam=\tencmcsc \scriptscriptfont\cmcscfam=\tencmcsc
\def\cmcsc{\fam\cmcscfam\tencmcsc} \textfont\cmcscfam=\tencmcsc \fi
\catcode`@=11
\newskip\ttglue
\gdef\tenpoint{\def\rm{\fam0\tenrm}
  \textfont0=\tenrm \scriptfont0=\sevenrm \scriptscriptfont0=\fiverm
  \textfont1=\teni \scriptfont1=\seveni \scriptscriptfont1=\fivei
  \textfont2=\tensy \scriptfont2=\sevensy \scriptscriptfont2=\fivesy
  \textfont3=\tenex \scriptfont3=\tenex \scriptscriptfont3=\tenex
  \def\mcal{\fam2 \tensy}  \def\mmit{\fam1 \teni}
  \textfont\itfam=\tenit \def\it{\fam\itfam\tenit}
  \textfont\slfam=\tensl \def\sl{\fam\slfam\tensl}
  \textfont\ttfam=\tentt \scriptfont\ttfam=\eighttt
  \scriptscriptfont\ttfam=\eighttt  \def\tt{\fam\ttfam\tentt}
  \textfont\bffam=\tenbf \scriptfont\bffam=\sevenbf
  \scriptscriptfont\bffam=\fivebf \def\bf{\fam\bffam\tenbf}
     \ifx\arisposta\amsrisposta    \ifnum\contaeuler=1
  \textfont\eufmfam=\teneufm \scriptfont\eufmfam=\seveneufm
  \scriptscriptfont\eufmfam=\fiveeufm \def\eufm{\fam\eufmfam\teneufm}
  \textfont\eufbfam=\teneufb \scriptfont\eufbfam=\seveneufb
  \scriptscriptfont\eufbfam=\fiveeufb \def\eufb{\fam\eufbfam\teneufb}
  \def\eurm{\teneurm} \def\eurb{\teneurb} \def\eusm{\teneusm}
  \def\eusb{\teneusb}    \fi    \ifnum\contaams=1
  \textfont\msamfam=\tenmsam \scriptfont\msamfam=\sevenmsam
  \scriptscriptfont\msamfam=\fivemsam \def\msam{\fam\msamfam\tenmsam}
  \textfont\msbmfam=\tenmsbm \scriptfont\msbmfam=\sevenmsbm
  \scriptscriptfont\msbmfam=\fivemsbm \def\msbm{\fam\msbmfam\tenmsbm}
     \fi      \ifnum\contacyrill=1     \def\cyrill{\tenwncyr}
  \def\cyrilb{\tenwncyb}  \def\cyrili{\tenwncyi}         \fi
  \textfont3=\tenex \scriptfont3=\sevenex \scriptscriptfont3=\sevenex
  \def\cmmib{\fam\cmmibfam\tencmmib} \scriptfont\cmmibfam=\sevencmmib
  \textfont\cmmibfam=\tencmmib  \scriptscriptfont\cmmibfam=\fivecmmib
  \def\cmbsy{\fam\cmbsyfam\tencmbsy} \scriptfont\cmbsyfam=\sevencmbsy
  \textfont\cmbsyfam=\tencmbsy  \scriptscriptfont\cmbsyfam=\fivecmbsy
  \def\cmcsc{\fam\cmcscfam\tencmcsc} \scriptfont\cmcscfam=\eightcmcsc
  \textfont\cmcscfam=\tencmcsc \scriptscriptfont\cmcscfam=\eightcmcsc
     \fi            \tt \ttglue=.5em plus.25em minus.15em
  \normalbaselineskip=12pt
  \setbox\strutbox=\hbox{\vrule height8.5pt depth3.5pt width0pt}
  \let\sc=\eightrm \let\big=\tenbig   \normalbaselines
  \baselineskip=\infralinea  \rm}
\gdef\ninepoint{\def\rm{\fam0\ninerm}
  \textfont0=\ninerm \scriptfont0=\sixrm \scriptscriptfont0=\fiverm
  \textfont1=\ninei \scriptfont1=\sixi \scriptscriptfont1=\fivei
  \textfont2=\ninesy \scriptfont2=\sixsy \scriptscriptfont2=\fivesy
  \textfont3=\tenex \scriptfont3=\tenex \scriptscriptfont3=\tenex
  \def\mcal{\fam2 \ninesy}  \def\mmit{\fam1 \ninei}
  \textfont\itfam=\nineit \def\it{\fam\itfam\nineit}
  \textfont\slfam=\ninesl \def\sl{\fam\slfam\ninesl}
  \textfont\ttfam=\ninett \scriptfont\ttfam=\eighttt
  \scriptscriptfont\ttfam=\eighttt \def\tt{\fam\ttfam\ninett}
  \textfont\bffam=\ninebf \scriptfont\bffam=\sixbf
  \scriptscriptfont\bffam=\fivebf \def\bf{\fam\bffam\ninebf}
     \ifx\arisposta\amsrisposta  \ifnum\contaeuler=1
  \textfont\eufmfam=\nineeufm \scriptfont\eufmfam=\sixeufm
  \scriptscriptfont\eufmfam=\fiveeufm \def\eufm{\fam\eufmfam\nineeufm}
  \textfont\eufbfam=\nineeufb \scriptfont\eufbfam=\sixeufb
  \scriptscriptfont\eufbfam=\fiveeufb \def\eufb{\fam\eufbfam\nineeufb}
  \def\eurm{\nineeurm} \def\eurb{\nineeurb} \def\eusm{\nineeusm}
  \def\eusb{\nineeusb}     \fi   \ifnum\contaams=1
  \textfont\msamfam=\ninemsam \scriptfont\msamfam=\sixmsam
  \scriptscriptfont\msamfam=\fivemsam \def\msam{\fam\msamfam\ninemsam}
  \textfont\msbmfam=\ninemsbm \scriptfont\msbmfam=\sixmsbm
  \scriptscriptfont\msbmfam=\fivemsbm \def\msbm{\fam\msbmfam\ninemsbm}
     \fi       \ifnum\contacyrill=1     \def\cyrill{\ninewncyr}
  \def\cyrilb{\ninewncyb}  \def\cyrili{\ninewncyi}         \fi
  \textfont3=\nineex \scriptfont3=\sevenex \scriptscriptfont3=\sevenex
  \def\cmmib{\fam\cmmibfam\ninecmmib}  \textfont\cmmibfam=\ninecmmib
  \scriptfont\cmmibfam=\sixcmmib \scriptscriptfont\cmmibfam=\fivecmmib
  \def\cmbsy{\fam\cmbsyfam\ninecmbsy}  \textfont\cmbsyfam=\ninecmbsy
  \scriptfont\cmbsyfam=\sixcmbsy \scriptscriptfont\cmbsyfam=\fivecmbsy
  \def\cmcsc{\fam\cmcscfam\ninecmcsc} \scriptfont\cmcscfam=\eightcmcsc
  \textfont\cmcscfam=\ninecmcsc \scriptscriptfont\cmcscfam=\eightcmcsc
     \fi            \tt \ttglue=.5em plus.25em minus.15em
  \normalbaselineskip=11pt
  \setbox\strutbox=\hbox{\vrule height8pt depth3pt width0pt}
  \let\sc=\sevenrm \let\big=\ninebig \normalbaselines\rm}
\gdef\eightpoint{\def\rm{\fam0\eightrm}
  \textfont0=\eightrm \scriptfont0=\sixrm \scriptscriptfont0=\fiverm
  \textfont1=\eighti \scriptfont1=\sixi \scriptscriptfont1=\fivei
  \textfont2=\eightsy \scriptfont2=\sixsy \scriptscriptfont2=\fivesy
  \textfont3=\tenex \scriptfont3=\tenex \scriptscriptfont3=\tenex
  \def\mcal{\fam2 \eightsy}  \def\mmit{\fam1 \eighti}
  \textfont\itfam=\eightit \def\it{\fam\itfam\eightit}
  \textfont\slfam=\eightsl \def\sl{\fam\slfam\eightsl}
  \textfont\ttfam=\eighttt \scriptfont\ttfam=\eighttt
  \scriptscriptfont\ttfam=\eighttt \def\tt{\fam\ttfam\eighttt}
  \textfont\bffam=\eightbf \scriptfont\bffam=\sixbf
  \scriptscriptfont\bffam=\fivebf \def\bf{\fam\bffam\eightbf}
     \ifx\arisposta\amsrisposta   \ifnum\contaeuler=1
  \textfont\eufmfam=\eighteufm \scriptfont\eufmfam=\sixeufm
  \scriptscriptfont\eufmfam=\fiveeufm \def\eufm{\fam\eufmfam\eighteufm}
  \textfont\eufbfam=\eighteufb \scriptfont\eufbfam=\sixeufb
  \scriptscriptfont\eufbfam=\fiveeufb \def\eufb{\fam\eufbfam\eighteufb}
  \def\eurm{\eighteurm} \def\eurb{\eighteurb} \def\eusm{\eighteusm}
  \def\eusb{\eighteusb}       \fi    \ifnum\contaams=1
  \textfont\msamfam=\eightmsam \scriptfont\msamfam=\sixmsam
  \scriptscriptfont\msamfam=\fivemsam \def\msam{\fam\msamfam\eightmsam}
  \textfont\msbmfam=\eightmsbm \scriptfont\msbmfam=\sixmsbm
  \scriptscriptfont\msbmfam=\fivemsbm \def\msbm{\fam\msbmfam\eightmsbm}
     \fi       \ifnum\contacyrill=1     \def\cyrill{\eightwncyr}
  \def\cyrilb{\eightwncyb}  \def\cyrili{\eightwncyi}         \fi
  \textfont3=\eightex \scriptfont3=\sevenex \scriptscriptfont3=\sevenex
  \def\cmmib{\fam\cmmibfam\eightcmmib}  \textfont\cmmibfam=\eightcmmib
  \scriptfont\cmmibfam=\sixcmmib \scriptscriptfont\cmmibfam=\fivecmmib
  \def\cmbsy{\fam\cmbsyfam\eightcmbsy}  \textfont\cmbsyfam=\eightcmbsy
  \scriptfont\cmbsyfam=\sixcmbsy \scriptscriptfont\cmbsyfam=\fivecmbsy
  \def\cmcsc{\fam\cmcscfam\eightcmcsc} \scriptfont\cmcscfam=\eightcmcsc
  \textfont\cmcscfam=\eightcmcsc \scriptscriptfont\cmcscfam=\eightcmcsc
     \fi             \tt \ttglue=.5em plus.25em minus.15em
  \normalbaselineskip=9pt
  \setbox\strutbox=\hbox{\vrule height7pt depth2pt width0pt}
  \let\sc=\sixrm \let\big=\eightbig \normalbaselines\rm }
\gdef\tenbig#1{{\hbox{$\left#1\vbox to8.5pt{}\right.\n@space$}}}
\gdef\ninebig#1{{\hbox{$\textfont0=\tenrm\textfont2=\tensy
   \left#1\vbox to7.25pt{}\right.\n@space$}}}
\gdef\eightbig#1{{\hbox{$\textfont0=\ninerm\textfont2=\ninesy
   \left#1\vbox to6.5pt{}\right.\n@space$}}}
\def\alternativefont#1#2{\ifx\arisposta\amsrisposta \relax \else
\xdef#1{#2} \fi}
\global\contaeuler=0 \global\contacyrill=0 \global\contaams=0
%
%
%
%
\newbox\fotlinebb \newbox\hedlinebb \newbox\leftcolumn
\gdef\makeheadline{\vbox to 0pt{\vskip-22.5pt
     \fullline{\vbox to8.5pt{}\the\headline}\vss}\nointerlineskip}
\gdef\makehedlinebb{\vbox to 0pt{\vskip-22.5pt
     \fullline{\vbox to8.5pt{}\copy\hedlinebb\hfil
     \line{\hfill\the\headline\hfill}}\vss} \nointerlineskip}
\gdef\makefootline{\baselineskip=24pt \fullline{\the\footline}}
\gdef\makefotlinebb{\baselineskip=24pt
    \fullline{\copy\fotlinebb\hfil\line{\hfill\the\footline\hfill}}}
\gdef\doubleformat{\shipout\vbox{\Landspec\makehedlinebb
     \fullline{\box\leftcolumn\hfil\columnbox}\makefotlinebb}
     \advancepageno}
\gdef\columnbox{\leftline{\pagebody}}
\gdef\line#1{\hbox to\hsize{\hskip\leftskip#1\hskip\rightskip}}
\gdef\fullline#1{\hbox to\fullhsize{\hskip\leftskip{#1}%
\hskip\rightskip}}
\gdef\footnote#1{\let\@sf=\empty
         \ifhmode\edef\#sf{\spacefactor=\the\spacefactor}\/\fi
         #1\@sf\vfootnote{#1}}
\gdef\vfootnote#1{\insert\footins\bgroup
         \ifnum\dimnota=1  \eightpoint\fi
         \ifnum\dimnota=2  \ninepoint\fi
         \ifnum\dimnota=0  \tenpoint\fi
         \interlinepenalty=\interfootnotelinepenalty
         \splittopskip=\ht\strutbox
         \splitmaxdepth=\dp\strutbox \floatingpenalty=20000
         \leftskip=\oldssposta \rightskip=\olddsposta
         \spaceskip=0pt \xspaceskip=0pt
         \ifnum\sinnota=0   \textindent{#1}\fi
         \ifnum\sinnota=1   \item{#1}\fi
         \footstrut\futurelet\next\fo@t}
\gdef\fo@t{\ifcat\bgroup\noexpand\next \let\next\f@@t
             \else\let\next\f@t\fi \next}
\gdef\f@@t{\bgroup\aftergroup\@foot\let\next}
\gdef\f@t#1{#1\@foot} \gdef\@foot{\strut\egroup}
\gdef\footstrut{\vbox to\splittopskip{}}
\skip\footins=\bigskipamount
\count\footins=1000  \dimen\footins=8in
\catcode`@=12
\tenpoint
\ifnum\unoduecol=1 \hsize=\tothsize   \fullhsize=\tothsize \fi
\ifnum\unoduecol=2 \hsize=\collhsize  \fullhsize=\tothsize \fi
\global\let\lrcol=L      \ifnum\unoduecol=1
\output{\plainoutput{\ifnum\tipbnota=2 \clearnmbnota\fi}} \fi
\ifnum\unoduecol=2 \output{\if L\lrcol
     \global\setbox\leftcolumn=\columnbox
     \global\setbox\fotlinebb=\line{\hfill\the\footline\hfill}
     \global\setbox\hedlinebb=\line{\hfill\the\headline\hfill}
     \advancepageno  \global\let\lrcol=R
     \else  \doubleformat \global\let\lrcol=L \fi
     \ifnum\outputpenalty>-20000 \else\dosupereject\fi
     \ifnum\tipbnota=2\clearnmbnota\fi }\fi
\def\ifdoublepage{\ifnum\unoduecol=2 }
\gdef\yespagenumbers{\footline={\hss\tenrm\folio\hss}}
\gdef\ciao{ \ifnum\fdefcontre=1 \endfdef\fi
     \par\vfill\supereject \ifnum\unoduecol=2
     \if R\lrcol  \headline={}\nopagenumbers\null\vfill\eject
     \fi\fi \end}

\newskip\olddsposta \newskip\oldssposta
\global\oldssposta=\leftskip \global\olddsposta=\rightskip

\def\filldots{\leaders\hbox to 1em{\hss.\hss}\hfill}
\def\inquadrb#1 {\vbox {\hrule  \hbox{\vrule \vbox {\vskip .2cm
    \hbox {\ #1\ } \vskip .2cm } \vrule  }  \hrule} }
 \def\newline{\hfil\break}
\def\jump{\vskip\baselineskip} \newskip\iinnffrr
\def\sjump{\iinnffrr=\baselineskip
          \divide\iinnffrr by 2 \vskip\iinnffrr}
\def\bjump{\vskip\baselineskip \vskip\baselineskip}
\newcount\nmbnota  \def\clearnmbnota{\global\nmbnota=0}
\newcount\tipbnota \def\letterfootnote{\global\tipbnota=1}

\def\note#1{\global\advance\nmbnota by 1 \ifnum\tipbnota=1
    \footnote{$^{\rm\nttlett}$}{#1} \else {\ifnum\tipbnota=2
    \footnote{$^{\nttsymb}$}{#1}
    \else\footnote{$^{\the\nmbnota}$}{#1}\fi}\fi}
\def\nttlett{\ifcase\nmbnota \or a\or b\or c\or d\or e\or f\or
g\or h\or i\or j\or k\or l\or m\or n\or o\or p\or q\or r\or
s\or t\or u\or v\or w\or y\or x\or z\fi}
\def\nttsymb{\ifcase\nmbnota \or\dag\or\sharp\or\ddag\or\star\or
\natural\or\flat\or\clubsuit\or\diamondsuit\or\heartsuit
\or\spadesuit\fi}   \clearnmbnota
\def\numberfootnote{\global\tipbnota=0} \numberfootnote
\def\setnote#1{\expandafter\xdef\csname#1\endcsname{
\ifnum\tipbnota=1 {\rm\nttlett} \else {\ifnum\tipbnota=2
{\nttsymb} \else \the\nmbnota\fi}\fi} }
\newcount\nbmfig  \def\clearnbmfig{\global\nbmfig=0}
\gdef\figure{\global\advance\nbmfig by 1
      {\rm fig. \the\nbmfig}}   \clearnbmfig
\def\setfig#1{\expandafter\xdef\csname#1\endcsname{fig. \the\nbmfig}}
 \def\endformula{\eqno\numero $$}
 \def\efr{\endformula}
\newcount\frmcount \def\clearfrmcount{\global\frmcount=0}
\def\numero{\global\advance\frmcount by 1   \ifnum\indappcount=0
  {\ifnum\cpcount <1 {\hbox{\rm (\the\frmcount )}}  \else
  {\hbox{\rm (\the\cpcount .\the\frmcount )}} \fi}  \else
  {\hbox{\rm (\applett .\the\frmcount )}} \fi}
\def\nameformula#1{\global\advance\frmcount by 1%
\ifnum\draftnum=0  {\ifnum\indappcount=0%
{\ifnum\cpcount<1\xdef\spzzttrra{(\the\frmcount )}%
\else\xdef\spzzttrra{(\the\cpcount .\the\frmcount )}\fi}%
\else\xdef\spzzttrra{(\applett .\the\frmcount )}\fi}%
\else\xdef\spzzttrra{(#1)}\fi%
\expandafter\xdef\csname#1\endcsname{\spzzttrra}
\eqno \hbox{\rm\spzzttrra} $$}
\def\nfr{\nameformula}    
\def\nameali#1{\global\advance\frmcount by 1%
\ifnum\draftnum=0  {\ifnum\indappcount=0%
{\ifnum\cpcount<1\xdef\spzzttrra{(\the\frmcount )}%
\else\xdef\spzzttrra{(\the\cpcount .\the\frmcount )}\fi}%
\else\xdef\spzzttrra{(\applett .\the\frmcount )}\fi}%
\else\xdef\spzzttrra{(#1)}\fi%
\expandafter\xdef\csname#1\endcsname{\spzzttrra}
  \hbox{\rm\spzzttrra} }      \clearfrmcount
\newcount\cpcount \def\clearcpcount{\global\cpcount=0}
\newcount\subcpcount \def\clearsubcpcount{\global\subcpcount=0}
\newcount\appcount \def\clearappcount{\global\appcount=0}
\newcount\indappcount \def\clearindappcount{\indappcount=0}
\newcount\sottoparcount 

\def\applett{\ifcase\appcount  \or {A}\or {B}\or {C}\or
{D}\or {E}\or {F}\or {G}\or {H}\or {I}\or {J}\or {K}\or {L}\or
{M}\or {N}\or {O}\or {P}\or {Q}\or {R}\or {S}\or {T}\or {U}\or
{V}\or {W}\or {X}\or {Y}\or {Z}\fi    \ifnum\appcount<0
\immediate\write16 {Panda ERROR - Appendix: counter "appcount"
out of range}\fi  \ifnum\appcount>26  \immediate\write16 {Panda
ERROR - Appendix: counter "appcount" out of range}\fi}
\clearappcount  \clearindappcount \newcount\connttrre
\def\clearconnttrre{\global\connttrre=0} \newcount\countref
\def\clearcountref{\global\countref=0} \clearcountref
\def\chapter#1{\global\advance\cpcount by 1 \clearfrmcount
                 \goodbreak\null\vbox{\jump\nobreak
                 \clearsubcpcount\clearindappcount
                 \itemitem{\ttaarr\the\cpcount .\qquad}{\ttaarr #1}
                 \par\nobreak\jump\sjump}\nobreak}
\def\section#1{\global\advance\subcpcount by 1 \goodbreak\null
               \vbox{\sjump\nobreak\ifnum\indappcount=0
                 {\ifnum\cpcount=0 {\itemitem{\ppaarr
               .\the\subcpcount\quad\enskip\ }{\ppaarr #1}\par} \else
                 {\itemitem{\ppaarr\the\cpcount .\the\subcpcount\quad
                  \enskip\ }{\ppaarr #1} \par}  \fi}
                \else{\itemitem{\ppaarr\applett .\the\subcpcount\quad
                 \enskip\ }{\ppaarr #1}\par}\fi\nobreak\jump}\nobreak}
\clearsubcpcount
\def\appendix#1{\global\advance\appcount by 1 \clearfrmcount
                  \goodbreak\null\vbox{\jump\nobreak
                  \global\advance\indappcount by 1 \clearsubcpcount
          \itemitem{ }{\hskip-40pt\ttaarr Appendix\ #1}
             \nobreak\jump\sjump}\nobreak}
\clearappcount \clearindappcount
\def\references{\goodbreak\null\vbox{\jump\nobreak
   \itemitem{}{\ttaarr References} \nobreak\jump\sjump}\nobreak}

\clearcpcount\clearcountref

\def\setchap#1{\ifnum\indappcount=0{\ifnum\subcpcount=0%
\xdef\spzzttrra{\the\cpcount}%
\else\xdef\spzzttrra{\the\cpcount .\the\subcpcount}\fi}
\else{\ifnum\subcpcount=0 \xdef\spzzttrra{\applett}%
\else\xdef\spzzttrra{\applett .\the\subcpcount}\fi}\fi
\expandafter\xdef\csname#1\endcsname{\spzzttrra}}
\newcount\draftnum \newcount\ppora   \newcount\ppminuti
\global\ppora=\time   \global\ppminuti=\time
\global\divide\ppora by 60  \draftnum=\ppora
\multiply\draftnum by 60    \global\advance\ppminuti by -\draftnum
\def\droggi{\number\day /\number\month /\number\year\ \the\ppora
:\the\ppminuti}     \global\draftnum=0
\def\draftcomment#1{\ifnum\draftnum=0 \relax \else
{\ {\bf ***}\ #1\ {\bf ***}\ }\fi} 
%
%
\catcode`@=11
\gdef\Ref#1{\expandafter\ifx\csname @rrxx@#1\endcsname\relax%
{\global\advance\countref by 1    \ifnum\countref>200
\immediate\write16 {Panda ERROR - Ref: maximum number of references
exceeded}  \expandafter\xdef\csname @rrxx@#1\endcsname{0}\else
\expandafter\xdef\csname @rrxx@#1\endcsname{\the\countref}\fi}\fi
\ifnum\draftnum=0 \csname @rrxx@#1\endcsname \else#1\fi}
\gdef\beginref{\ifnum\draftnum=0  \gdef\Rref{\fairef}
\gdef\endref{\scriviref} \else\relax\fi
\ifx\risposta\mplarisposta \ninepoint \fi
\parskip 2pt plus.2pt \baselineskip=12pt}
\def\Reflab#1{[#1]} \gdef\Rref#1#2{\item{\Reflab{#1}}{#2}}
\gdef\endref{\relax}  \newcount\conttemp
\gdef\fairef#1#2{\expandafter\ifx\csname @rrxx@#1\endcsname\relax
{\global\conttemp=0 \immediate\write16 {Panda ERROR - Ref: reference
[#1] undefined}} \else
{\global\conttemp=\csname @rrxx@#1\endcsname } \fi
\global\advance\conttemp by 50  \global\setbox\conttemp=\hbox{#2} }
\gdef\scriviref{\clearconnttrre\conttemp=50
\loop\ifnum\connttrre<\countref \advance\conttemp by 1
\advance\connttrre by 1
\item{\Reflab{\the\connttrre}}{\unhcopy\conttemp} \repeat}
\clearcountref \clearconnttrre
\catcode`@=12
\ifx\risposta\mplarisposta \def\Reflab#1{#1.} \letterfootnote \fi

\def\slashchar#1{\setbox0=\hbox{$#1$} \dimen0=\wd0
     \setbox1=\hbox{/} \dimen1=\wd1 \ifdim\dimen0>\dimen1
      \rlap{\hbox to \dimen0{\hfil/\hfil}} #1 \else
      \rlap{\hbox to \dimen1{\hfil$#1$\hfil}} / \fi}
\ifx\oldchi\undefined \let\oldchi=\chi
  \def\cchi{{\raise 1pt\hbox{$\oldchi$}}} \let\chi=\cchi \fi

\def\frac#1#2{{\textstyle{#1 \over #2}}}

\def\half{\ifinner {\scriptstyle {1 \over 2}}\else {1 \over 2} \fi}

\def\simge{\rlap{\raise 2pt \hbox{$>$}}{\lower 2pt \hbox{$\sim$}}}
\def\simle{\rlap{\raise 2pt \hbox{$<$}}{\lower 2pt \hbox{$\sim$}}}

\def\vbig#1#2{{\vbigd@men=#2\divide\vbigd@men by 2%
\hbox{$\left#1\vbox to \vbigd@men{}\right.\n@space$}}}

%
%
\newcount\fdefcontre \newcount\fdefcount \newcount\indcount
\newread\filefdef  \newread\fileftmp  \newwrite\filefdef
\newwrite\fileftmp     \def\strip#1*.A {#1}
\def\futuredef#1{\beginfdef
\expandafter\ifx\csname#1\endcsname\relax%
{\immediate\write\fileftmp {#1*.A}
\immediate\write16 {Panda Warning - fdef: macro "#1" on page
\the\pageno \space undefined}
\ifnum\draftnum=0 \expandafter\xdef\csname#1\endcsname{(?)}
\else \expandafter\xdef\csname#1\endcsname{(#1)} \fi
\global\advance\fdefcount by 1}\fi   \csname#1\endcsname}

\def\beginfdef{\ifnum\fdefcontre=0
\immediate\openin\filefdef \jobname.fdef
\immediate\openout\fileftmp \jobname.ftmp
\global\fdefcontre=1  \ifeof\filefdef \immediate\write16 {Panda
WARNING - fdef: file \jobname.fdef not found, run TeX again}
\else \immediate\read\filefdef to\spzzttrra
\global\advance\fdefcount by \spzzttrra
\indcount=0      \loop\ifnum\indcount<\fdefcount
\advance\indcount by 1   \immediate\read\filefdef to\spezttrra
\immediate\read\filefdef to\sppzttrra
\edef\spzzttrra{\expandafter\strip\spezttrra}
\immediate\write\fileftmp {\spzzttrra *.A}
\expandafter\xdef\csname\spzzttrra\endcsname{\sppzttrra}
\repeat \fi \immediate\closein\filefdef \fi}
\def\endfdef{\immediate\closeout\fileftmp   \ifnum\fdefcount>0
\immediate\openin\fileftmp \jobname.ftmp
\immediate\openout\filefdef \jobname.fdef
\immediate\write\filefdef {\the\fdefcount}   \indcount=0
\loop\ifnum\indcount<\fdefcount    \advance\indcount by 1
\immediate\read\fileftmp to\spezttrra
\edef\spzzttrra{\expandafter\strip\spezttrra}
\immediate\write\filefdef{\spzzttrra *.A}
\edef\spezttrra{\string{\csname\spzzttrra\endcsname\string}}
\iwritel\filefdef{\spezttrra}
\repeat  \immediate\closein\fileftmp \immediate\closeout\filefdef
\immediate\write16 {Panda Warning - fdef: Label(s) may have changed,
re-run TeX to get them right}\fi}
\def\iwritel#1#2{\newlinechar=-1
{\newlinechar=`\ \immediate\write#1{#2}}\newlinechar=-1}
\global\fdefcontre=0 \global\fdefcount=0 \global\indcount=0
%
%
\null
%
%
%
%
\input psfig

%
\loadamsmath
\loadeuler
\mathchardef\bbalpha="710B
\mathchardef\bbbeta="710C
\mathchardef\bbgamma="710D
\mathchardef\bbdelta="710E
\mathchardef\bblambda="7115
\mathchardef\bbxi="7118
\mathchardef\bbpsi="7112
\mathchardef\bbrho="7116
\mathchardef\bbomega="7121
\mathchardef\sdir="2D6E
\mathchardef\dirs="2D6F

\def\bal{{\bfmath\bbalpha}}
\def\bb{{\bfmath\bbbeta}}
\def\bgamma{{\bfmath\bbgamma}}
\def\bd{{\bfmath\bbdelta}}

\def\ba{{\bfmath a}}

\def\bq{{\bfmath q}}
\def\bg{{\bfmath g}}
\def\bH{{\bfmath H}}

\def\SU{{\rm SU}}
\def\SO6R{{\rm SO}(6)_{R}}

\def\b1{{\bf 1}}
\def\ba{{\bfmath a}}

\def\balpha{{\bfmath\bbalpha}}

\pageno=0\baselineskip=14pt
\nopagenumbers{
\line{\hfill SWAT/97/145}
\line{\hfill\tt hep-th/9704011}
\line{\hfill March 1997}
\vfill
\centerline{\capsone Semi-classical Quantization in N=4 Supersymmetric}
\centerline{\capsone Yang-Mills Theory and Duality}
\bjump\sjump
\centerline{\scaps Christophe Fraser and Timothy J. Hollowood}
\sjump
\centerline{\sl Department of Physics, University of Wales Swansea,}
\centerline{\sl Singleton Park, Swansea SA2 8PP, U.K.}
\centerline{\tt c.fraser, t.hollowood @swansea.ac.uk}
\bjump\bjump

\centerline{\capsone ABSTRACT}\sjump\sjump
At a generic point in the moduli space of vacua of an
$N=4$ supersymmetric gauge theory with arbitrary gauge group
the Higgs force does not
cancel the magneto-static force between magnetic monopoles of distinct charge.
As a consequence the moduli space of magnetically charged solutions
is related in a simple way to those of the SU(2) theory. This leads to
a rather simple test of S-duality. On certain
subspaces of the moduli space of vacua the forces between distinct 
monopoles cancel and the test of S-duality becomes more complicated. 

\vfill\eject}

\yespagenumbers\pageno=1
\parskip5pt

In this letter, we re-assess the evidence for S-duality in the spectrum of BPS
states in $N=4$ 
supersymmetric Yang-Mills theories with arbitrary gauge groups. 
Our conclusion is that at a
generic point in the moduli space of vacua the spectrum of such states
can be deduced from the spectrum of the SU(2) theory 
[\Ref{SEN},\Ref{Evidence}], although we cannot prove that it is complete. 
The Higgs field in both $N=2$ and $N=4$ supersymmetric
gauge theories transforms as a vector of an internal global $R$-symmetry
SO($N_R$) and this complication will be seen to have profound
implications for the spectrum of BPS states. The modifications to 
the semi-classical quantization of monopoles caused by having a vector Higgs
field were considered in the context of an $N=2$ theory in
[\Ref{DECAY}]. In this letter we consider these modifications in a
theory with $N=4$ supersymmetry. 
The essential modification in the vector Higgs model can be 
stated quite simply as the fact that between distinct static 
monopoles, the scalar interaction does not generically cancel
the magneto-static interaction implying there are no
static BPS solutions consisting of well separated distinct
monopoles. 

In order to compute the force between two well-separated monopoles, we
shall follow the approach of [\Ref{GM},\Ref{LWY1}] and effectively treat them 
as point particles. The Higgs field is denoted $\Phi^I$,
where the SO($N_R$) vector index runs $I=1,\ldots,6$ in an $N=4$ theory.
The Higgs field of a single monopole behaves for large $r$ as 
[\Ref{DECAY}]
$$
\Phi^I_{\bal} = \ba^I\cdot\bH - {1\over er} 
\lambda^I_{\bal}(\bal^\star\cdot{\bH}),\qquad
\lambda^I_{\bal} = {\ba^I\cdot\bal \over
\parallel \ba^I\cdot\bal \parallel},
\nfr{DEFL}
where $\Phi^I_0=\ba^I \cdot {\bH}$ is the VEV of the Higgs field
lying in some Cartan subalgebra of the Lie algebra $g$
of the gauge group. This ensures that the Higgs VEV satisfies
$[\Phi^I_0,\Phi^J_0]=0$.
The vector $\bal$ is some positive root of the Lie algebra 
and by definition the co-root is $\bal^\star=\bal/\bal^2$. In the
above $\lambda_\bal^I$ is a unit $N_R$ vector and 
$\parallel X^I \parallel=\sqrt{X^IX^I}$ denotes the 
length of the vector $X^I$. The solution 
has an asymptotic magnetic field
$$
\vec B={\vec r\over er^3}\bal^\star\cdot\bH,
\efr
and so $\bg=\bal^\star$ is the vector magnetic charge of the monopole.
The mass of the monopole is 
$$
M={4\pi\over e}\parallel \ba^I\cdot\bal\parallel.
\efr
 
Consider the superposition of two such monopole solutions 
associated to positive roots $\bgamma$ and $\bd$. For large separation the
solution will be approximated by the linear superposition
$$
\Phi^I = \Phi_0^I + \Delta\Phi^I_{\bgamma}
+ \Delta\Phi^I_{\bd},
\efr
where $\Delta\Phi^I_{\bal}=\Phi^I_{\bal}-\Phi^I_0$. The
potential due to long-range Higgs field can be deduced by considering
the effect of the correction to the Higgs field by the second monopole
on the first. The effect of the second monopole is to change the Higgs
VEV at the first monopole to
$$
\ba^I\rightarrow\ba^I-{1\over er}\lambda^I_{\bd}\bd^\star.
\efr
Plugging this into the mass formula of the first monopole we deduce
for static solutions that the leading order term in the  
potential describing the scalar interaction is 
$$
V_{{\rm Higgs}} = -{4\pi\over e^2r} \bgamma^\star\cdot\bd^\star
\sum_{I=1}^{N_{{R}}}\lambda^I_{\bgamma}\lambda^I_{\bd}.
\efr
The magneto-static scalar potential is coulomb-like and involves the 
inner product of the two vector magnetic charges:
$$
V_{{\rm em}} = {4\pi\over e^2r} \bgamma^\star\cdot\bd^\star.
\efr
The total interaction is then
$$
V_{{\rm Higgs}}+V_{{\rm em}} = {4\pi\over e^2r}
\bgamma^\star\cdot\bd^\star
\left( 1 - \sum_{I=1}^{N_{R}}
\lambda^I_{\bgamma}\lambda^I_{\bd} \right).
\efr
It follows that when $\bgamma$ and $\bd$ are two
distinct positive roots the net force is
attractive if $\bgamma\cdot\bd<0$, and repulsive if $\bgamma\cdot\bd>0$.
If $\bgamma\cdot\bd\neq0$ the force cancels only 
when $\lambda^I_{\bgamma}=\lambda^I_{\bd}$.
When $\bgamma\cdot\bd = 0$ the two monopole do not interact at long
range but there can be short range forces when the
monopole cores overlap.
 
The consequences of this result are rather far-reaching. Spherically
symmetric monopole solutions can be constructed whose magnetic charge
is any co-root $\bal^\star$ of the Lie algebra. For theories with a
single real Higgs field, Weinberg showed [\Ref{EW1}] that when $\bal$
is a non-simple root (with respect to a fundamental Weyl chamber 
defined by the Higgs VEV) then the solution could be always deformed
away from spherical 
symmetry leading to an asymptotic configuration consisting of static 
constituent monopoles associated to the simple co-roots---the so-called
`fundamental' monopoles. In the single component 
Higgs model, the fact that these 
additional moduli appear is manifested in the fact that the overall long-range
force between the constituent monopoles is always zero. On the contrary, 
in the vector Higgs model, 
the forces do not generically cancel. This is an indication 
that the additional moduli describing the degrees-of-freedom 
corresponding to separating the monopole into its fundamental 
constituents are not present and the original monopole is stable, 
and thus is itself `fundamental'. 
 
To confirm this heuristic picture,
we will show that generically 
in the vector Higgs model, the moduli space of monopoles with 
magnetic charge $\bg=n\bal^\star$, where $\bal$ is {\it any\/} 
root of $g$ and $n$ is a positive integer, is
identical---up to a scale factor $1/\bal^2$---to 
the moduli space of $n$ SU(2) monopoles. 
As a consequence, at most points in the moduli space of vacua
GNO duality [\Ref{GNO}] is rather simple to test.
In a nutshell, 
GNO duality states that a strongly coupled theory with gauge group $G$ has 
an alternative description as a weakly coupled theory with gauge group 
$G^\star$---the group whose roots are the co-roots of
$G$---such that the role of gauge bosons and monopoles is
interchanged. 
The conjecture therefore leads to a prediction for the spectrum of
monopoles which can be tested at weak coupling using semi-classical
techniques.
Unlike [\Ref{GNO}], it is not necessary to consider the role of the 
centre of $G$ since all the fields are adjoint-valued and space-time is
Minkowskian. 
 
The GNO duality conjecture is extended in the presence of a
theta angle [\Ref{SEN},\Ref{SDUAL}].
This `S-duality' requires an infinite set 
of stable dyon states with quantum numbers 
$({\bg},{\bq})=(n_m\bal^\star,n_e\balpha)$, 
where $n_m$ and $n_e$ are co-prime integers. The quantum number $\bq$ 
determines the electric charge of the dyon state as we explain later.
This prediction is a generalization of the SU(2) case [\Ref{SEN}].  
As explained in [\Ref{SDUAL}], there is a simple way of 
determining whether the dyon $(n_m\bal^\star,n_e\balpha)$ 
corresponds under S-duality to a gauge boson with gauge group 
$G$ or $G^\star$. Evidence for this extended duality conjecture now
requires finding these additional dyon states---and no other ones---as 
stable bound states of monopoles.
Stable dyon bound states correspond to harmonic forms on the centred
monopole moduli space. For gauge group SU(2) the $n$-monopole moduli spaces 
have the form
$$
{\cal M}_n = {\Bbb R}^3\times{S^1\times\tilde{\cal M}_n^0\over{\Bbb Z}_n}.
\efr
The factors
$\tilde{\cal M}_n^0$ are not explicitly known for $n\geq 3$, but 
evidence that they exhibit the relevant harmonic forms was
found in [\Ref{SEN},\Ref{Evidence}].
 
In theories with larger gauge groups, we shall argue that 
the moduli space of a monopole with magnetic
charge $\bal^\star$ is generically
$$
{\cal M}_{\bal^\star}= {\Bbb R}^3 \times S^1, 
\efr
i.e. the one monopole moduli space of the SU(2) theory. More
generally, we have ${\cal M}_{n\bal^\star}\simeq{\cal M}_n$, up to a
possible rescaling.
However, on special subspaces in the space of vacua  
where certain monopoles reach the threshold for
decay into their components, there is a discontinuous change:
$$
{\cal M}_{\bal^\star}={\Bbb R}^3 \times S^1\rightarrow 
{\Bbb R}^3 \times {{\Bbb R}^1 \times {\cal M}_{{\rm rel}}
\over {\Bbb Z}}, 
\efr
to a higher dimensional space reflecting the fact that the monopole 
can dissociate into stable constituents.
This change is only discontinuous to first order in 
the moduli space approximation. To higher orders, 
one should describe the change by introducing a potential on 
${\cal M}_{{\rm rel}}$.

On these special subspaces, 
GNO duality requires a single bound state associated
to a unique normalisable harmonic form on ${\cal M}_{{\rm rel}}$. 
This programme has been carried out with great success in the case of 
SU$(n)$ gauge theory 
[\Ref{LWY1},\Ref{SU3},\Ref{Chalmers},\Ref{One}].
There has also been some analysis on subspaces when a 
non-abelian gauge symmetry is restored [\Ref{NonAbelian}].
 
The mass formula 
for BPS states in $N=4$ (or $N=2$) supersymmetric Yang-Mills theory can be
established by considering the energy:
$$ 
U={1\over2}\int d^3x\,{\rm Tr}\left(\vec E^2+\vec B^2+
\parallel D_0\Phi^I\parallel^2+\parallel\vec D\Phi^I\parallel^2
+\sum_{I<J}\left[\Phi^I,\Phi^J\right]^2\right),
\nfr{ENER}
where we have assumed that the fermion fields are zero.
This can be expressed in the following form:
$$\eqalign{
U={1\over2}\int d^3x\,{\rm Tr}&\left(\parallel\eta^I\vec E+
\rho^I\vec B-\vec D
\Phi^I\parallel^2+\parallel D_0\Phi^I\parallel^2+
\sum_{I<J}\left[\Phi^I,\Phi^J\right]^2\right.\cr
&\qquad\qquad\left.
+2\sum_{I=1}^{N_R}
\eta^I\vec E\cdot\vec D\Phi^I+2\sum_{I=1}^{N_R}
\rho^I\vec B\cdot\vec D\Phi^I\right),\cr}
\nfr{ENT}
where $\rho^I$ and $\eta^I$ are two orthonormal SO$(N_R)$ vectors.
From \ENT\ we deduce a bound for the energy of a configuration:
$$
U\geq\sum_{I=1}^{N_R}\left(\rho^IQ_M^I+\eta^IQ_E^I\right),
\nfr{UBOG}
where $Q^I_E$ and $Q^I_M$ are defined by Gauss's law
$$
Q^I_M = \int_{S^2_\infty}d\vec S\cdot
{\rm Tr} \left(\vec B \Phi^I \right), \qquad
Q^I_E = \int_{S^2_\infty}d\vec S\cdot 
{\rm Tr} \left(\vec E \Phi^I\right).
\nfr{DME}
Here the integrals are taken over the sphere at spatial infinity. 

The most stringent bound for the energy is achieved by maximizing the
right-hand-side of \UBOG\ as a function of $\rho^I$ and $\eta$
subject to the fact that they are orthonormal. The solution is that
$\rho^I$ and $\eta^I$ are in the plane defined by
$Q^I_M$ and $Q^I_E$ as illustrated in Fig.~1. Here $\alpha$ is the
angle between $Q^I_M$ and $Q^I_E$ and 
$$
\tan\theta={\parallel Q^I_E\parallel\cos\alpha\over
\parallel Q^I_M\parallel+\parallel
Q^I_E\parallel\sin\alpha}.
\efr
This gives the Bogomol'nyi Bound for the masses of particles of 
given electric and magnetic charges:
$$
M^2 \geq \parallel Q^I_M\parallel^2+\parallel
Q^I_E\parallel^2+2\parallel Q^I_M\parallel\parallel
Q^I_E\parallel\sin\alpha.
\nfr{BOUND}
A configuration which saturates the
bound, i.e.~a BPS configuration, must satisfy the equations
$$
D_0\Phi^I=0,\qquad \eta^I\vec E+
\rho^I\vec B=\vec D\Phi^I,\qquad[\Phi^I,\Phi^J]=0.
\nfr{BOGE}

\midinsert{\bjump
\centerline{
\psfig{figure=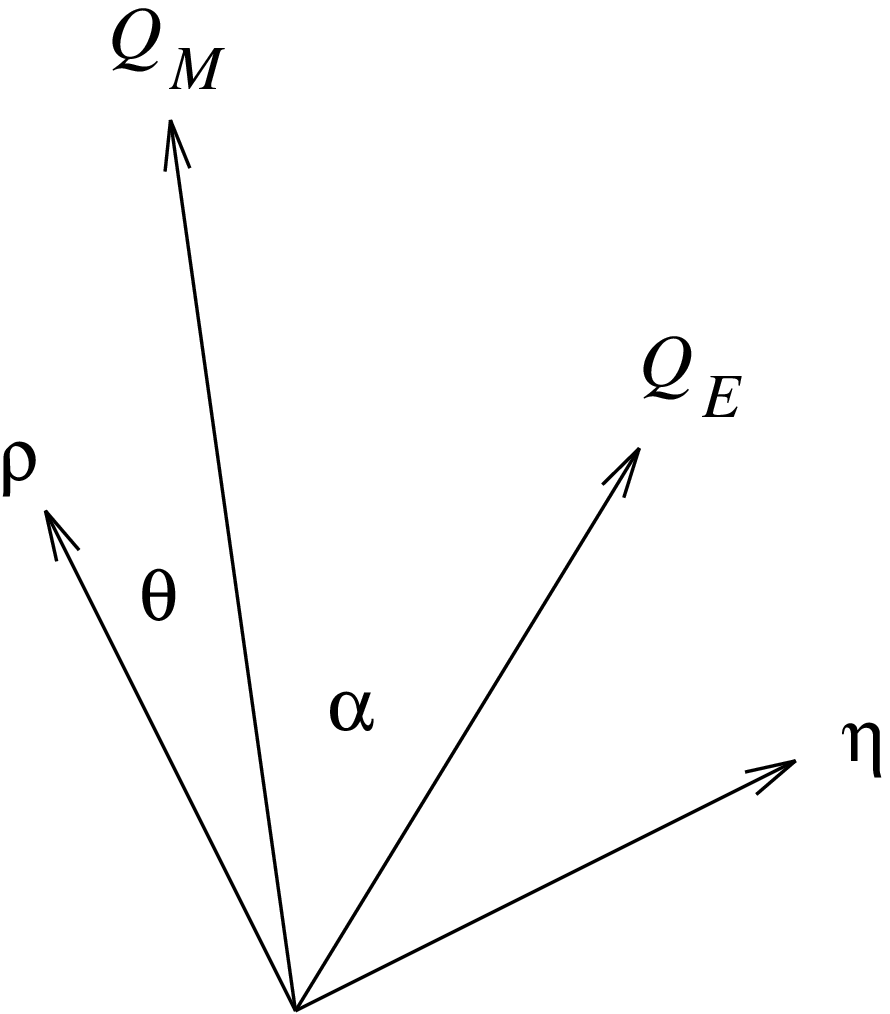,height=6cm}}
\bjump\bjump
\centerline{Figure 1. The relation between $\rho^I$, $\eta^I$,
$Q_M^I$ and $Q_E^I$}
\sjump
}
\endinsert

The mass of a BPS state matches precisely what one expects from the $N=4$
superalgbera. This has two central charges $z_\pm$ 
expressed in terms of the electric and
magnetic charges:
$$
z_\pm^2= \parallel Q^I_M\parallel^2+\parallel
Q^I_E\parallel^2\pm 2\parallel Q^I_M\parallel\parallel
Q^I_E\parallel\sin\alpha.
\nfr{CENC}
The mass of any state has to be greater than or equal to the greater
of the two central charges,
i.e.~$z_+$ with our definitions. There are two types of
BPS state. Firstly with $M=z_+>z_-$, which implies that
$\alpha>0$, i.e.~$Q^I_M\not\propto
Q^I_E$. Such states transform in $2^6$ dimensional 
representations of $N=4$ supersymmetry rather than the canonical $2^8$
dimensional representation. If $\alpha=0$, i.e.~the
electric and magnetic charge vectors are parallel $Q^I_M\propto
Q^I_E$, then a BPS state with $M=z_\pm$ transforms in a $2^4$
dimensional (ultra short) representation of $N=4$ supersymmetry. In
this case since $\alpha=0$ these states have a mass
$$
M^2=\parallel Q^I_M\parallel^2+\parallel Q^I_E\parallel^2.
\nfr{PBB}
In this case the vectors $\rho^I$ and $\eta^I$ are only determined
up to an SO$(N_R)$ rotation around the common axis of $Q^I_M$ and $Q^I_E$. 

At spatial infinity the Higgs field is equal to its VEV
$\Phi_0^I=\ba^I\cdot\bH$, and we can define the vector electric and magnetic
charges via
$$
Q^I_M = {4\pi\over e} \ba^I\cdot\bg, \qquad
Q^I_E = e \ba^I\cdot\tilde\bq.
\nfr{VECC}
With this normalization
the vector magnetic charge \bg\ can be any vector of the co-root
lattice of $g$. The generalized Dirac quantization 
condition, taking account the existence of a theta angle [\Ref{Witten}],
leads to a quantization of vector electric charge:
$$
\tilde\bq = \bq + {\theta \over 2\pi} \bg,
\efr
where $\bq$ is any vector of the root lattice of $g$. We choose to
label BPS states with the vector quantum numbers $(\bg,\bq)$.

The only known solutions of the Bogomol'nyi equations \BOGE\ are those
that follow from embeddings of the 
SU(2) dyon solutions with a single real Higgs field and these
solutions automatically have $Q^I_M\propto Q^I_E$, i.e.~$\bq\propto\bg$.
In [\Ref{SDUAL}], it was shown that a minimal set of states required 
by the conjectured form of S-duality have vector charges $(\bg,\bq)=
(n_m\bal^{\star},n_e\bal)$, 
where $\bal$ is a root of $g$ and $n_m$ and $n_e$ are two co-prime
integers. These states certainly have $\bq\propto\bg$ and so 
$Q_E^I\propto Q_M^I$. They have a mass 
$$
M=e \left\vert n_e + \tau {n_m\over\bal^2} \right\vert
\parallel \ba^I\cdot \bal \parallel,
\nfr{MASS}
and transform in ultra-short representations of $N=4$
supersymmetry. In the above $\tau = {\theta/2\pi} + {4\pi i/e^2}$.
The middle-dimensional BPS multiplets with $2^6$ states
contain particles with spin $>1$ and would presumably 
complicate the picture as far as duality is concerned and although
there are no
known solutions to the Bogomol'nyi equations with $Q_E^I\not\propto Q_M^I$
we cannot rule out their existence.

The corresponding situation in $N=2$ supersymmetric gauge theories is
rather different. In that case the supersymmetry algebra only has one
central charge and there is only one type of BPS state which can have
$Q_M^I$ parallel to $Q_E^I$, or otherwise. In fact there {\it are\/} 
BPS states in these theories which do have $Q_E^I\not\propto Q_M^I$ 
arising from quantum corrections to the electric
charge of a dyon [\Ref{FH},\Ref{ME}].
This does not occur in the $N=4$ theory because the quantum corrections 
vanish exactly.
 
One may wonder why the BPS states only have a vector magnetic charge 
$\bg=n_m\bal^\star$, for some co-root $\bal^\star$, 
rather than {\it any\/} vector of the co-root
lattice. At the present stage of understanding this is a conjecture
which remains to be proved. In the semi-classical approximation one
would first have to construct the moduli space of solutions to the
Bogomol'nyi equations \BOGE\ with a given magnetic charge and then
argue that only when $\bg=n_m\bal^\star$ is
there a bound-state corresponding to a stable BPS state.
Some progress has been made in
the construction of moduli spaces corresponding to 
solutions with magnetic charge which are not a multiple of a co-root
in the SU$(n)$ theory [\Ref{Chalmers}]. 
However these solutions exist in a theory with a single real Higgs
field and they cannot be embedded in the vector Higgs
model in any obvious way.

The moduli space of vacua ${\cal M}_{\rm
vac}$ is parameterized by $\ba^I$ modulo reflections by the Weyl group
of $G$. We can use this freedom to fix $\sum_I\sigma^I\bal^I$ in the
dominant Weyl chamber where $\sigma^I$ is some arbitrary fixed
vector. The vector $\sum_I\sigma^I\bal^I$ then defines a set of
simple roots $\bal_i$, $i=1,\ldots,{\rm rank}(g)$, and fixes a notion
of positive and negative roots. 
From the mass formula \MASS\ it follows that BPS states are at the
threshold for decay into other BPS states on 
certain submanifolds of ${\cal M}_{\rm vac}$. There are two
possible types of decay. The first is familiar from the SU(2) theory
where a state $(n_m\bal^\star,n_e\bal)$ with $n_m$ and $n_e$ having a common
factor $p$, $n_m=pn'_m$ and $n_e=pn'_e$, is at the threshold for decay into 
$p$ states of charge $(n'_m\bal^{\star},n'_e\bal)$. Kinematically, these
decays can occur at any point in ${\cal M}_{\rm vac}$.  
When the charges of a dyon are proportional to a non-simple root 
$\bal$ a second kind of decay can occur. In that case
there exist pairs of positive roots $\bgamma$ and $\bd$ such that
$$
\bal^\star=N\bgamma^\star+M\bd^\star,
\nfr{DECA} 
for positive integers $N$ and 
$M$, and the dyon can decay
into a number of dyons of magnetic charge $\bgamma^\star$ and
$\bd^\star$ [\Ref{DECAY}].
The important point is that kinematically 
this decay can only occur on a certain subspace of ${\cal M}_{\rm
vac}$ of co-dimension $5$ defined by the condition
$\lambda^I_\bgamma=\lambda^I_\bd$
which is precisely the condition that the force between the
$\bgamma^\star$ and $\bd^\star$ dyons vanishes.
This defines the ``Curve of Marginal Stability'' (CMS) denoted 
$C_{\bgamma,\bd}$. 
There are cases in the non-simply-laced
groups when $\bgamma\cdot\bd=0$
and there is no long-range force between the $\bgamma^\star$ 
and the $\bd^\star$ dyons at any point in ${\cal M}_{\rm
vac}$. Nevertheless, the $\bal^\star$ dyon is not generically at threshold
for decay into the $\bgamma^\star$ and $\bd^\star$ dyon. 
The resolution of this paradox is that there are generically 
no solutions of the Bogomol'nyi equations corresponding to separated 
$\bgamma^\star$ and $\bd^\star$ dyons and the moduli space of BPS solutions
of charge $\bal^\star$ is simply isomorphic to ${\cal M}_1$ 
describing the spherically symmetric $\bal^\star$ dyon.
 
In order to find explicit solutions to the Bogomol'nyi equations
\BOGE\ consider the analogous equations for the theory with a
single real Higgs field $\phi$ and gauge group SU(2):
$$
D_0\phi=0,\qquad\vec E=\left(\sin\xi\right)\vec D\phi,\qquad\vec
B=\left(\cos\xi\right)\vec D\phi,
\nfr{BOGR}
where $\tan\xi=Q_E/Q_M$
and $Q_M$ and $Q_E$ are defined as in \DME\ with $\Phi^I$ replaced by $\phi$.
These solutions have $Q_M=(4\pi v/e)n_m$, where $n_m$ is
an integer and $v$ is the VEV of $\phi$.
The electric charge is subject to the Dirac
quantization condition:
$$
Q_E=ev\left(n_e+{\theta\over2\pi}n_m\right),\qquad{n_e}\in{\Bbb Z}.
\efr 

In order to write down solutions to \BOGE\ we pick a regular embedding of the
Lie algebra $su(2)$ in the Lie algebra of the gauge group
associated to a positive root $\bal$ and
defined by the three generators $\{E_{\pm\bal},\bal^\star\cdot\bH\}$. We
denote by $\phi^\bal$ and $A^\bal_\mu$ the solution of \BOGR\ embedded
in the theory with a larger gauge group 
using this $su(2)$ Lie subalgebra. The ansatz for the solution of the
full Bogomol'nyi equations \BOGE\ is 
$$
\Phi^I=
\lambda_\bal^I\phi^\bal+\left(\ba^I-(\ba^I\cdot\bal^\star)
\bal\right)\cdot\bH,\qquad A_\mu=A^\bal_\mu,
\nfr{ANSATZ}
where the VEV of the SU(2) Higgs field is 
$v=\parallel\ba^I\cdot\bal\parallel$ 
and $\lambda_\bal^I$ is defined in \DEFL. This ensures that the
embedded solution \ANSATZ\ has the correct VEV. The solution has
$(\bg,\bq)=(n_m\bal^\star,n_e\bal)$ and in particular $Q_M^I\propto Q_E^I$.

Using this construction we can find purely magnetically charged 
solutions with vector magnetic charge $\bg=n_m\bal^\star$.
The moduli space of these solutions can be probed locally
by finding the zero-modes for fluctuations in the Higgs and gauge
field. Equivalently, it is more convenient to 
find the Dirac zero-modes since
they are paired with the bosonic modes by unbroken 
supersymmetries [\Ref{Blum}].

In the appendix, we show that the zero-mode equation can be cast
in a form familiar from the real Higgs theory [\Ref{EW1}]. It is
convenient to define two orthogonal projections of $\Phi^I$ 
with respect to the  vector $\lambda_\bal^I$:
$$
\Phi=\sum_{I=1}^{N_R}\lambda_{\bal}^I\Phi^I \quad \hbox{and} \quad 
\hat\Phi^I=\Phi^I-\lambda_{\bal}^I\sum_{J=1}^{N_R}\lambda_{\bal}^J\Phi^J.
\nfr{DPROJ}
$\Phi$ is then Weinberg's ansatz for the monopole solutions
in the real Higgs model [\Ref{EW1}]
with VEV $\sum_I\lambda_{\bal}^I\ba^I$. 
The orthogonal projection is a constant:
$$
\hat\Phi^I=\hat\ba^I\cdot\bH=
\left(\ba^I-\lambda_{\bal}^I\sum_{J=1}^{N_R}\lambda_{\bal}^J\ba^J
\right)\cdot\bH.
\efr

We are now in a position to describe the zero modes. 
The idea is to consider the relation between the
corresponding zero modes for the real Higgs
theory [\Ref{EW1}]. One first expands the
adjoint-valued modes in a Cartan-Weyl basis.
There are four zero-modes taking values in the $su(2)$ Lie algebra
$\{E_{\pm\bal},\bal^\star\cdot\bH\}$ corresponding to overall 
translations and charge rotations. The remaining generators $E_\bb$ can be
grouped into representations under the embedding SU(2) labelled by
the total isospin $t$. The member $E_\bb$ with the lowest value of
$t_3=\bal\cdot\bb/\bal^2=-t$ can be used to label each isospin multiplet.
The zero modes associated to each pair of isospin multiplets
containing $E_{\pm\bb}$ describe the
freedom for the dyon to decay as in \DECA\ with $\bgamma$ and $\bd$
being expressed in terms of $\bal$ and $\bb$ [\Ref{DECAY}]. 
This means that the definition of the CMS $C_{\bgamma,\bd}$ is equivalent to
$\hat\ba^I\cdot\bb=0$, $\forall I$.
The computation in the appendix shows that these zero-modes of the
real Higgs theory are lifted in the vector Higgs theory unless they 
are annihilated by $\hat\Phi^I$.
It follows immediately that at a generic point in the moduli space which
is not on a CMS
all zero-modes, except those corresponding to translations and charge
rotations, are lifted. Hence generically the spherically
symmetric monopole with $\bg=\bal^\star$ will be 
stable. In the case of higher magnetic charges
the zero-modes associated to the root $\bal$ can be identified
with the zero-modes of the SU(2) $n_m$-monopole solution since
they all take values in the embedding $su(2)$.
This proves that the monopole moduli space of 
charge $\bg=n_m\bal^\star$ is identical to the SU(2) 
charge $n_m$ monopole moduli space up to a scale factor 
$1/\bal^2$ from the normalization of the killing form. 
After quantization the spectrum of
stable BPS dyons have charges $(\bg,\bq)=(n_m\bal^\star,n_e\bal)$
where $(n_m,n_e)$ are co-prime integers.

\newline

\noindent TJH is supported by a PPARC Advanced Fellowship.
CF would like to thank David Olive and David Tong for useful conversations.

\appendix{}

In this appendix we study the Dirac equation in the background
of the classical fields \ANSATZ\ and show how to
relate the zero-modes of the single Higgs to the vector Higgs model.
A convenient way to obtain the $d=4$ $N=4$ Dirac
equation is by dimensional reduction from the $d=10$ $N=1$ 
Dirac equation. We follow closely the conventions of
Osborn [\Ref{OS}], with minor notational changes.
The $d=4$ $N=4$ supermultiplet of adjoint valued fields 
contains, as well as the gauge field and the 
six scalars $\Phi^I$, four Majorana fermions 
$\psi^{a}_{\alpha}$, where $a=1,\ldots,4$
denotes the $\SO6R$ (or more correctly $\SU (4)$) spinor index, 
and $\alpha$ denotes the usual Dirac spinor index. 

The Dirac equation obtained by dimensionally reducing
$i\Gamma^AD_A\psi=0$ is
$$
\left\{ i\gamma^\mu D_\mu
-i\alpha^m\Phi^{m}
+\gamma_5\beta^{\dot{n}}\Phi^{3+\dot{n}} \right\}
\psi = 0,
\nfr{DIRAC}
where $m,\dot{n}=1,2,3$ and the gauge and Higgs fields act by adjoint
action on $\psi$. The $\gamma$-matrices carry space-time 
spinor indices,
while the $4\times 4$ matrices 
$\alpha^m$ and $\beta^{\dot{n}}$
carry $\SO6R$ spinor indices. They are 
antisymmetric, (anti-)self-dual matrices with real entries 
satisfying
$$\eqalign{
\lbrace\alpha^m,\alpha^n\rbrace = -2\delta^{mn} \qquad &\qquad
\lbrace\beta^{\dot{m}},\beta^{\dot{n}}\rbrace 
= -2\delta^{\dot{m}\dot{n}}\cr
\left[\alpha^m,\alpha^n\right] = -2\epsilon^{mnp}\alpha^p \quad &\quad
\left[\beta^{\dot{m}},\beta^{\dot{n}}\right] = -2
\epsilon^{\dot{m}\dot{n}\dot{p}}\beta^{\dot{p}}\cr
\left[\alpha^m,\beta^{\dot{n}}\right] &= 0. \cr}
\efr
In a static background we look for stationary solutions 
$\psi({\vec r},t) = e^{-iEt} \psi({\vec r})$, leading to the
the Dirac Hamiltonian equation
$$
\left\{
i\gamma^0\gamma_i D_i
-i\gamma^0\alpha^m \Phi^m
-\gamma^0\gamma_5\beta^{\dot{n}}
\Phi^{3+\dot{n}} \right\} \psi = E \psi. 
\nfr{HAMIL}
Squaring this we find
$$
\left\{
-D_i^2+\parallel\Phi^I\parallel^2
+{1\over 2}\gamma_{ij}F_{ij}
-\left(\gamma_i\lambda^m\alpha^m
+i\gamma_i\gamma_5\lambda^{3+\dot{n}}\beta^{\dot{n}}
\right)B_i\right\} \psi=E^2 \psi,
\efr
where $\gamma_{ij}=i\epsilon_{ijk}\gamma_0\gamma_5\gamma_k$. Now 
introduce the following set of Euclidean $\gamma$-matrices:
$$\eqalign{
\tilde\gamma_i &= \gamma_0\gamma_i,\qquad
\tilde\gamma_4 = \gamma_0\lambda^m\alpha^m
+i\gamma_0\gamma_5\lambda^{3+\dot{n}}\beta^{\dot{n}}\cr
\tilde\gamma_5 &= \tilde\gamma_1\tilde\gamma_2\tilde\gamma_3\tilde\gamma_4 
= -i\gamma_0\gamma_5\lambda^m\alpha^m
+\gamma_0\lambda^{3+\dot{n}}\beta^{\dot{n}}.\cr}
\efr
This gives
$$
\left\{-D_i^2+\parallel\Phi^I\parallel^2 -i \gamma_5
\tilde\gamma_iB_i\left(1+\tilde\gamma_5\right)\right\}\psi=E^2\psi
\nfr{POLP}
Define the chiral projections 
$\psi_{\pm}={1\over 2} \left\{1 \pm\tilde\gamma_5\right\}\psi$ 
and a representation of the Pauli matrices $\sigma_i=\gamma_5 \tilde\gamma_i$.
Using this \POLP\ becomes
$$
\left(-D_i^2+\parallel\Phi^I\parallel^2-
2i\vec\sigma\cdot\vec B\right) \psi_{+}=E^2\psi_+,\qquad
\left(-D_i^2+\parallel\Phi^I\parallel^2\right) \psi_{-}=E^2\psi_- .
\nfr{INTN}
Introducing the projections of $\Phi^I$ defined in
\DPROJ: $\parallel\Phi^I\parallel^2= 
\Phi^2+\parallel\hat\Phi^I\parallel^2$,
and defining
$$
{\cal D}=-i\vec\sigma\cdot\vec D-i\Phi,\qquad
{\cal D}^*=-i\vec\sigma\cdot\vec D+i\Phi,
\efr
we have
$$
{\cal D}^*{\cal D}=-D_i^2+\Phi^2-2i\vec\sigma\cdot\vec B,\qquad
{\cal D}{\cal D}^*=-D_i^2+\Phi^2.
\efr
In terms of these operators \INTN\ becomes
$$
\left({\cal D}^*{\cal D} + 
\parallel\hat\Phi^I\parallel^2\right)
\psi_{+}=E^2\psi_+,\qquad
\left({\cal D}{\cal D}^* + 
\parallel\hat\Phi^I\parallel^2\right)
\psi_{-}=E^2\psi_-.
\efr
${\cal D}{\cal D}^*$ is a positive definite operator and has no non-trivial 
zero-modes. The number of zero-modes of ${\cal D}^*{\cal D}$
is determined in [\Ref{EW1}]. These will only survive as zero-modes of
\HAMIL\ if annihilated by $\hat\Phi^I$. 

\references

\beginref
\Rref{Blum}{J. Gauntlett, Nucl. Phys. {\bf B411} (1994) 443\newline
J. Blum, Phys. Lett. {\bf B333} (1994) 92-97}
\Rref{DECAY}{T. Hollowood, `Semi-Classical Decay of Monopoles
in N=2 Gauge Theory' {\tt hep-th/9611106}}
\Rref{GM}{N. Manton,
Phys. Lett. {\bf B110} (1982) 54\newline  N. Manton and G. Gibbons, 
Phys. Lett. {B356} (1995) 32, {\tt hep-th/9506052} } 
\Rref{LWY1}{K. Lee, E.J. Weinberg and P. Yi, 
Phys. Rev. {\bf D54} (1996) 1633, {\tt hep-th/9602167}}
\Rref{GNO}{P. Goddard, J. Nuyts and D. Olive,
Nucl. Phys. {\bf B125} (1977) 1}
\Rref{SDUAL}{N. Dorey, C.Fraser, T.J. Hollowood and M.A.C. Kneipp,
 Phys. Lett. {\bf B383} (1996) 422, {\tt hep-th/9605069 }} 
\Rref{SEN}{A. Sen, Phys. Lett. {\bf B329} (1994) 217, {\tt hep-th/9402032}}
\Rref{Evidence}{M. Porrati, 
Phys. Lett. {\bf B377} (1996) 67, {\tt hep-th/9505187}\newline
G. Segal and A. Selby, Commun. Math. Phys. {\bf 177} (1996) 775\newline
F. Ferrari, `The dyon spectra of finite gauge theories',
{\tt hep-th/9702166}}
\Rref{SU3}{S.A. Connell, `The Dynamics of the ${\rm SU}(3)$ $(1,1)$
Magnetic Monopole', unpublished preprint available by anonymous
ftp from : \newline
{\tt <ftp://maths.adelaide.edu.au/pure/- mmurray/oneone.tex>},\newline
K. Lee, E.J. Weinberg and P. Yi, 
Phys. Lett. {\bf B376} (1996) 97, {\tt hep-th/9601097}\newline
J.P. Gauntlett and D. A. Lowe, 
Nucl. Phys. {\bf B472} (1996) 194, {\tt hep-th/9601085}}
\Rref{One}{G.W. Gibbons and P. Rychenkova, `Hyperk\"ahler Quotient
Construction of BPS Monopole Moduli Spaces', {\tt hep-th/9608085},\newline
M.K. Murray, `A Note on the $(1,1,\dots ,1)$ Monopole Metric', 
{\tt hep-th/9605054},\newline
G.W. Gibbons, Phys. Lett. {\bf B382} (1996) 53, 
{\tt hep-th/9603176}}
\Rref{Chalmers}{G. Chalmers, `Multi-Monopole Moduli 
Spaces for SU(n) Gauge Group', {\tt hep-th/9605182}} 
\Rref{NonAbelian}{A.S. Dancer, Nonlinearity {\bf 5} (1992) 1355\newline
N. Dorey, C.Fraser, T.J. Hollowood and M.A.C. Kneipp, 
`Non-Abelian Duality in N=4 Supersymmetric Gauge theories',
{\tt hep-th/9512116},\newline
K. Lee, E.J. Weinberg and P. Yi, 
Phys. Rev. {\bf D54} (1996) 6351, {\tt hep-th/9605229}}
\Rref{OS}{H. Osborn, Phys. Lett. {\bf B83} (1979) 321}
\Rref{Witten}{E. Witten, Phys. Lett. {\bf B86} (1979) 283}
\Rref{FH}{C. Fraser and T.J. Hollowood, `On the Weak Coupling
Spectrum of N=2 Supersymmetric SU(n) Gauge Theory', {\sl to appear
in\/}: Nucl. Phys. {\bf B}, {\tt hep-th/9610142}}
\Rref{ME}{T.J. Hollowood, in preparation} 
\Rref{EW1}{E.J. Weinberg, Phys. Rev. {\bf D20} (1979) 936;
Nucl. Phys. {\bf B167} (1980) 500; Nucl. Phys. {\bf B203} (1982) 445}
\endref

\ciao